# THE FORECASTING OF 3G MARKET IN INDIA BASED ON REVISED TECHNOLOGY ACCEPTANCE MODEL

Dr. Sudha Singh[1]  Dr. D. K. Singh[2] , Dr. M. K. Singh[3]  and Sujeet Kumar Singh[4]

[1]Associate Professor, PG Department of Computer Science and Engg., Bengal College of Engineering And Technology, Bidhan Nagar, Durgapur, West Bengal-713212(INDIA).
*Sudha_2k6@yahoo.com*

[2] Head, Department of Electronics and Telecommunications & IT, BIT Sindri, Dhanbad, Jharkhand-828123(INDIA).
*dksingh_bit@yahoo.com*

[3] Professor and Head, PG Department of Mathematics, Ranchi University,Ranchi, Jharkhand-834008(INDIA).
*mithileshkumarsingh@gmail.com*

[4]Application Developer, JPMorgan Chase, Andheri,Mumbai, INDIA.
*sujeetkumarsingh@gmail.com*

## ABSTRACT

*3G, processor of 2G services, is a family of standards for mobile telecommunications defined by the International Telecommunication Union [1]. 3G services include wide-area wireless voice telephone, video calls, and wireless data, all in a mobile environment. It allows simultaneous use of speech and data services and higher data rates.3G is defined to facilitate growth, increased bandwidth and support more diverse applications. The focus of this study is to examine the factors affecting the adoption of 3G services among Indian people. The study adopts the revised Technology Acceptance Model by adding five antecedents-perceived risks, cost of adoption, perceived service quality, subjective norms, and perceived lack of knowledge. Data have collected from more than 400 school/college/Institution students & employees of various Government/Private sectors using interviews & various convenience sampling procedures and analyzed using MS excel and MATLAB. Result shows that perceived usefulness has the most significant influence on attitude towards using 3G services, which is consistent with prior studies. Of the five antecedents, perceived risk and cost of adoption are found to be significantly influencing attitude towards use. The outcome of this study would be beneficial to private and public telecommunication organizations, various service providers, business community, banking services and people of India. Research findings and suggestions for future research are also discussed.*

## I. INTRODUCTION

3G, processor of 2G services, is a family of standards for mobile telecommunications defined by the International Telecommunication Union[1].3G services include wide-area wireless voice telephone, video calls, and wireless data, all in a mobile environment. It allows simultaneous use of speech and data services and higher data rates.3G is defined to facilitate growth, increased bandwidth and support more diverse applications. These applications are mainly made possible due to the enhanced data rates as a result of the 2-8MBPS bandwidth availabilities. Some of the applications are (1) Mobile TV - Due to the high data transfer rate being offered due to 3G, TV can be viewed on Mobile Phones. For this have to tie up with a service provider, through which the content can be accessed. For example Apalaya for BSNL(India); (2) Video Conferencing - One can conduct a video conferencing through mobile using the available network. All this is feasible due to the enhanced bandwidth of 2 MBPS; (3) Tele-medicine - This is an extended feature of video conferencing where a remote person can be given attention by a doctor located at a distant place (4) Location Based Services - These are some services which we can access on the dependence of the service provider. Some of the examples are weather updates, live road traffic view, vehicle tracking etc; (5) Video on Demand - People can view videos on demand from their service provider. For providing this service the service provider should have collaborations with content providers For example Perceptknorigin (India). This is again possible due to high buffering speed possible due to the 3G network.

India's population is around 1.17 billion people and 72.2 % of the population lives in villages. Population under poverty line is 22%. Population under 0-14 , 15-64 and 65+ years are 30.8%, 64.3 % and 4.9% respectively[24].  Literacy rate is 61%.In India one out of 14000 is going for higher education. Working population of India is 699.9 million. At the end of 2009, the total number of mobile users in India will be more than 500 million [24]. In India, there are two telecom operators who provide 3G service at present-MTNL and BSNL. MTNL is targeting four field jumps in its 3G mobile subscriber base to 6 laks by March 2010 in Mumbai and Delhi. Also, BSNL launched 3G service on Feb.27 in 70 cities of India[24]. Initially launch of 3G services in India was welcomed by every mobile phone enthusiast. But immediately security concerns were raised about 3G services. For example, data transfers and voice calls were encoded but it can not be decoded on real time basis. Security agencies were serious about it and working on it. The adoption of 3G in all over India is growing slowly but steadily[24]. We know that 3G services have already become popular in Japan, UK, Hong Kong, Australia, Sweden and Denmark etc.

## II. LITERATURE REVIEW

To examine the factors affecting the adoption process of new Information Communication Technology(ICTs), the Technology Acceptance Model[5] has been most widely cited in the literature.Building on the Theory of Reasoned Action(TRA), originally developed by Fishbein and Ajzein's[6], TAM aimed to examine the attitude and belief of users that affect their acceptance or rejection of adopting ICTs. TAM has been used in the study of software, World Wide Web(Lederer et al.[12],Van D H[22]), Intranet(Horton et al.[10]),E-commerce(Pavlou[18]),Technology Training(Marler[16]),e-shopping (Liaoetal[13],Gefen[8] and Forsythe[7]), M-commerce(Nysveen et al.[17], Yang[23],Ching et al[4]), Internet banking service(Polatoglv[20], Chan et al.[3],Litter D. et al[14],Jin et al[11]), Mobile banking(Luarn et al[15]), Mobile Internet service(Pedersen[19]) etc.

## III.  RESEARCH MODEL

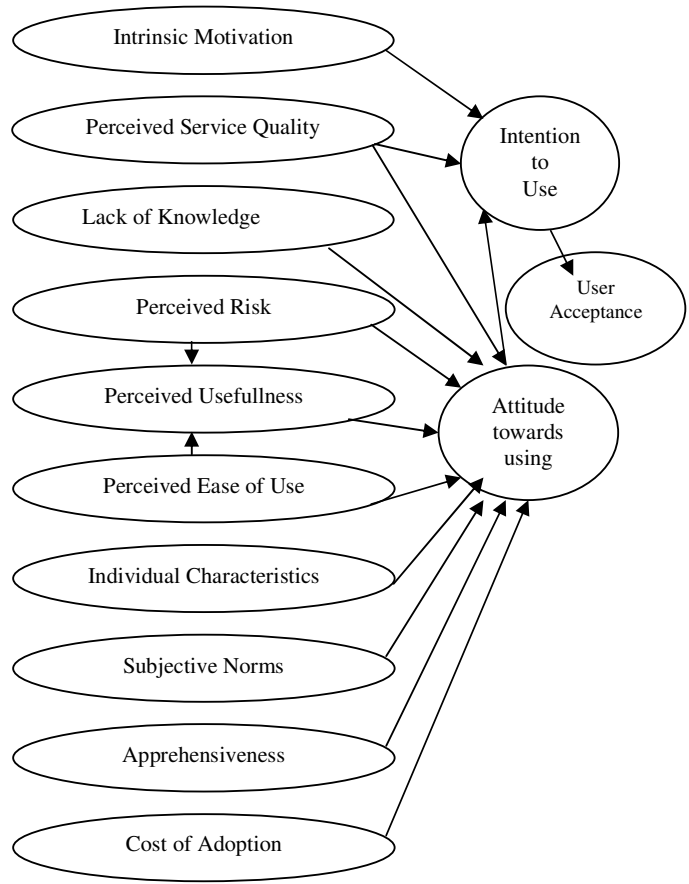

Figure 1: Research Model

The following hypotheses are postulated:

(1) Perceived usefulness (PU) Positively influences Attitude Towards Using(ATU) 3G services.

(2) Perceived Ease of Use(PEU) positively influences Attitude Towards Using(ATU) 3G services.

(3) Perceived Ease of Use(PEU) positively influences Perceived Usefulness of 3G services.

(4) Perceived Risk (PR) negatively influences Attitude Towards Using(ATU) 3G services.

(5) Perceived Risk (PR) negatively influences Perceived Usefulness of 3G services.

(6) Lack of Knowledge (PLK) negatively influences Attitude Towards Using(ATU) 3G services.

(7) Cost of Adoption(CA)  negatively influences Attitude Towards Using(ATU) 3G services

(8) Subjective Norms (SN) positively influences Attitude Towards Using(ATU) 3G services.

(9) Intrinsic Motivation (IM) positively influences Attitude Towards Using(ATU) 3G services.

(10) Perceived Service Quality (PSQ) negatively influences Attitude Towards Using(ATU) 3G services.

(11) Perceived Service Quality (PSQ) positively influences Intension to Use 3G services.

(12) Apprehensiveness negatively influences Attitude Towards Using (ATU) 3G services.

(13) Individual characteristics(IC) positively influence Attitude Towards Using (ATU) 3G services.

This study is based on a modified version of TAM by adding SN, CA, LK, PSQ, and PR as the independent variable. The model modification and selection of variables are based on scenario of India and the fact that 3G is still at its early stage in India.

## IV.  RESULTS AND CONCLUSION

This study was conducted in major cities of India using Google survey/Emails/Questionnaires/Interviews and other sampling procedures. The data were collected from students of School/College/Institutions, employees of Government/Private sector and highly educated housewives who are using 3G services. Significant data (400) were analyzed using MS Excel and MATLAB. College students from metros and educated housewives are more interested to adopt 3g services because of their exposure, educational level and income potential.

The ten construct of interest to this study were IM, PSQ, LK, PR, PU, PEU, IC, SN, A and CA. All construct, except for SN, IM, IC and A were measured using multiple item perceptual scales. All constructs were adopted from revalidated instruments of prior research wherever possible, and rephrased to relate specifically to the context of 3G service adoption. Wherever possible, respondents were asked to assess a total of 20 items by indicating their agreement with a set of statements on a five-point Likert scale, ranging from 1(strongly agree) to 5(strongly disagree).

Among the 400 respondents, 65% were male and 45% were female. Initial Information is given below:

| Item | 1 | 2 | 3 | 4 | 5 |
|------|---|---|---|---|---|
| Perceived usefulness | 12.5 | 60 | 25 | 2.5 | 0 |
| Perceived risk | 62.5 | 37.5 | 0 | 0 | 0 |
| Perceived ease of use | 37.5 | 50 | 12.5 | 0 | 0 |
| Perceived service quality | 25 | 50 | 25 | 0 | 0 |
| Individual characteristics | 50 | 25 | 25 | 0 | 0 |
| cost of adoption | 50 | 37.5 | 12.5 | 0 | 0 |
| apprehensiveness | 50 | 25 | 25 | 0 | 0 |
| subjective norms | 0 | 50 | 37.5 | 12.5 | 0 |
| lack of knowledge | 12.5 | 37.5 | 25 | 12.5 | 12.5 |
| Intrinsic Motivation | 37.5 | 37.5 | 25 | 0 | 0 |

The measures of ATU, IM, PU, PEU, IC and A were adopted from previous studies related to TAM, mainly from the studies of Davis[5] and yang[23]. These six construct were measured using six items. The measure of SN was adopted from the study of Fishbein and Ajzen[6] by using a single item on family influence. The measure of PR were adopted from the study of Chan and Lu[3] by using two items, focusing only on two types of risk-privacy and safety. The measures of CA were adopted from the study of Hung et al[8] by using two items. This construct focused on the current charge and extra cost that the consumers are willing to pay for the 3G services. Consistent with prior research, we also collected demographic information such as gender, caste etc[4].

From the collected information we found that mean deviation of the study variables PU, PR, PEU PSQ, IC, A, SN, LK and IM were 2.5, 1.5, 2, 2, 2, 1.25, 2.75, 2.25, 3 and 2.25 respectively. Reliability was assessed using internal consistency scores, calculated by the composite reliably scores.

The results from the hypothesis testing are:

(1) PU is the most important factor that influences ATU.

(2) IM is significantly influencing Intention to Use(ITU).

(3) PSQ is significantly influencing ATU.

(4) A and IC are the other factors that influences ATU. But it has less significance.

(5) Even though PEU directly influences ATU, the indirect influence through PU is even greater. Thus, PU is a strong mediator between PEU and ATU. It is consistent to results of ching[3].

(6) CA and PR are the other two important factors which significantly influencing ( in negative) ATU.

(7) SN and LK has no significant influence on ATU.

The result shows that there is a promising future for 3G-services in India. Perceived risk and cost of adoption are great concerns.  The outcome of this study would be beneficial to private and public telecommunication organizations, various service providers, business community, banking services and people of India.

The results of the study must be used with caution, as the sample size is relatively small compared to number of mobile users in India. Further research should be conducted for secured 3G services. Alternative models such as the triangulation framework proposed by Anckar and D'Incau[2] and perceived value model by pura[21] may be considered in the future research to make comparison as which model may best represent the Indian consumer in relation to the adoption of 3G services[4].

*Acknowledgement*: The authors would like to thank A. K. Sharma, Ankit, Amit and Rakesh for data collection.

## AUTHORS

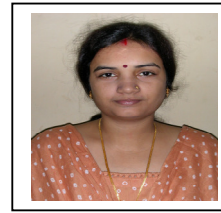

**Dr. Sudha Singh** received her M.Tech. degree in computer science from Birla Institute of Technology,Mesra,Ranchi,India, in 2002and was placed First Class First. She obtained her Ph.D. degree in Computer Science and Engineering from BIT Sindri,(VBU),India. She has more than 14 publications. Her research interest is in recent advances in Data communication, network security, coding theory, signal processing and combinatorics. sudha_2k6@yahoo.com

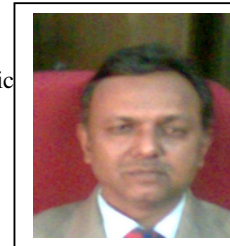

**Dr. Dharmendra K Singh** is presently working as Head, Department of Electronic and Communication & Information Technology, BIT Sindri, Dhanbad. He has more than 20 years of teaching experience. He is heading the department of Electronics and Communication & Information technology since 2002. He is instrumental in starting the curriculum on information technology. He has published more than 30 papers in journals and conferences. He has already supervised 01 thesis at doctoral level in computer Science & Engg and 05 research scholars are presently enrolled under him for their doctoral degree. The area of research he works are Coding theory, cryptography, optical Amplifiers, Photonic Crystal Fibers, e-Governance and technical education Planning. dksingh_bit@yahoo.com

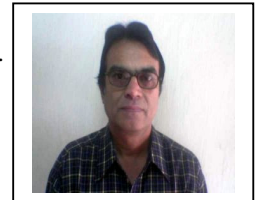

**Dr. Mithilesh Kumar Singh** is University Professor at Ranchi University Ranchi. He is actively involved in research in various disciplines which includes Coding theory and Network Security, Mathematical modelling, Discrete and Combinatorial Mathematics and Mathematical Biology. He has *Published* more than 45 papers in International and National Journals. He has already supervised 05 thesis at doctoral level in Applied Mathematics and Computer Science and 05 research scholars are presently enrolled under him for their doctoral degree. He has also visited IISc, Bangalore and JNU, New Delhi as scientist during the period 1996-98. He is successfully promoting students for higher education and research from 1975. Presently he is Director MCA Program and Head of PG Department of Mathematics at Ranchi University, Ranchi. mithileshkumarsingh@gmail.com

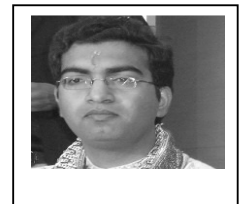

**Sujeet Kumar Singh** has done his Graduation in Computer Applications from BIT Mesra, Ranchi. He has done his Masters in Computer Applications from Cochin University of Science And Technology. After that he joined Tata Consultancy Services Limited as Assistant System Engineer in 2006. He was part of many successful projects there and worked with many clients like ABN AMRO and Mckinsey. He joined JPMorgan Chase in December 2009 and currently working there as Application Developer. He enjoys solving difficult problems and studying current and future generation of wireless technology.